\documentclass[twocolumn]{revtex4}
\usepackage{graphicx,epsf}
\usepackage{color}
\usepackage{dcolumn}
\usepackage{bm}
\usepackage{amsfonts}
\usepackage{mathrsfs}

\usepackage{subfig}

\begin{document}

\title{Short Wavelength Geodesic Acoustic Mode Excitation by Energetic Particles}

\author{Liu Chen$^{1, 2}$, Zhiyong Qiu$^{1}$  and Fulvio Zonca$^{3, 1}$}

\affiliation{$^1$Institute for    Fusion Theory and Simulation and Department of Physics, Zhejiang University, Hangzhou, P.R.C\\
$^2$Department of   Physics and Astronomy,  University of California, Irvine CA 92697-4575, U.S.A.\\
$^3$ ENEA, Fusion and Nuclear Safety Department,
C. R. Frascati, Via E. Fermi 45, 00044 Frascati (Roma), Italy}

\begin{abstract}
Taking the collisionless damping of geodesic acoustic mode (GAM) as an example, the  physics processes underlying  wave particle resonances in the short wavelength limit  are clarified. As illustrative application, GAM excitation by energetic particles  in short wavelength limit is investigated assuming a single pitch angle slowing-down fast ion equilibrium distribution function. Conditions for  this energetic particle-induced GAM (EGAM) to be unstable are discussed. 
\end{abstract}

\maketitle

Recently, due to geodesic acoustic mode (GAM)  \cite{NWinsorPoF1968} excitation by energetic particles (EPs) in the large drift orbit limit \cite{MSasakiPoP2016}, there has been renewed interest in   wave-particle resonances  at short wavelength, which was firstly investigated in Ref. \citenum{FZoncaEPL2008}  for the collisionless damping of GAM, and  presented   later providing detailed derivation and physics interpretation \cite{ZQiuPPCF2009}. The same approach was also applied to study   quasi-linear transport of EPs by drift wave turbulence \cite{WZhangPRL2008,ZFengPoP2013}. However,  the understanding  of the underlying physics processes   proposed in recent literature, e.g. \cite{MSasakiPoP2016}, may yield to some mis-interpretation and inconsistency with the existing  theoretical framework. In this brief communication, our aim is to clarify the underlying physics processes for wave-particle resonance in the short wavelength limit  and,  as illustrative application,  investigate EP-induced GAM (EGAM) \cite{RNazikianPRL2008,GFuPRL2008,ZQiuPPCF2010} excitation by fast ions with large magnetic drift orbits.

To discuss the physics picture of wave-particle resonance in the short wavelength limit, we take electrostatic GAM  collisionless damping originally discussed in \cite{FZoncaEPL2008,ZQiuPPCF2009} as example.   For the clarity of discussion, we  assume small but finite electron temperature, i.e., $\tau\equiv T_e/T_i\ll1$ such that $|\widetilde{\delta\phi}_G/\overline{\delta\phi}_G|\sim \tau k_r\rho_{ti}\ll1$ while one still has $|\omega_{tr,e}|\gg|\omega_G|$. Here, $\widetilde{\delta\phi}_G $  and  $\overline{\delta\phi}_G$ are respectively the $m\neq0$ and $m=0$  components of the perturbed scalar potential;  $\omega_{tr}\equiv v_{\parallel}/(qR_0)$ is the transit frequency,  $k_r$ is the radial wavenumber and $\rho_{ti}$ is the ion Larmor radius at thermal velocity. In this limit,  consistent with the  short wavelength assumption of interest here, the perturbed electron response  (distribution function) to GAM is $\delta f_e=0$, and the GAM dispersion relation   can be derived from the quasi-neutrality condition:
\begin{eqnarray}
\sum_{s}\langle \delta f_s\rangle=0.\label{eq:QN}
\end{eqnarray}
Here, $\langle\cdots\rangle$ denotes velocity space integration, subscript $s$ denotes different ions species and, thus,  equation (\ref{eq:QN}) can also be applied to study  EGAM excitation by EPs. $\delta f_s$ can be expressed as $\delta f_s=e\partial_EF_0\delta\phi/m+\exp[i(m_ic)/(eB^2)\mathbf{k}\times\mathbf{B}\cdot\mathbf{v}]\delta H$, and  the nonadiabatic response can be   derived from the following linear gyrokinetic equation \cite{PRutherfordPoF1968,JTaylorPP1968}:
\begin{eqnarray}
\left(-i\omega+\omega_{tr}\partial_{\theta}+i\omega_d\right)\delta H_k=i\omega(e/m_i)\partial_E F_0 J_k\overline{\delta\phi}_G,\label{eq:LinearGKE}
\end{eqnarray}
with   $\omega_d=\hat{\omega}_d\sin\theta=k_r\rho_{ti}v_{ti}(v^2_{\perp}/2+v^2_{\parallel})/(v^2_{ti}R_0)\sin\theta$ being the magnetic drift frequency due to geodesic curvature, $J_k\equiv J_0(k_r\rho_{ti})$ with $J_0$ being Bessel function of zero-order accounting for  finite Larmor radius effects, $v_{ti}\equiv \sqrt{2T_i/m_i}$ being the ion thermal velocity, $E=v^2/2$ and other notations are stardard.

Noting that $\omega_G\simeq v_{ti}/R_0\sim q\omega_{tr,i}\gg\omega_{b,i}\simeq \sqrt{\epsilon}\omega_{tr,i}$,  and assuming well circulating particles in the large aspect ratio limit, equation (\ref{eq:LinearGKE})  can be solved and yields, for $v_{\parallel}>0$, 
\begin{eqnarray}
\delta H_s=\frac{\omega}{\omega_{tr}}\hat{S}e^{-\psi(\theta)}\int^{\theta}_{-\infty} e^{\psi(\theta')}d\theta'.\label{eq:deltaH_general}
\end{eqnarray} 
Here, $\hat{S}\equiv -i(e/m_i)\partial_E F_0J_G\overline{\delta\phi}_G$, $\psi(\theta)\equiv -i(\omega\theta+\hat{\omega}_d\cos\theta)/\omega_{tr}$, $\omega_b$ is the bounce frequency of trapped particles, and $\epsilon\equiv r/R_0$ is the inverse aspect ratio.  Similar expression can also be obtained for $v_{\parallel}<0$.

Noting that 
\begin{eqnarray}
e^{i\hat{\Lambda}cos\theta}=\sum_l i^lJ_l(\hat{\Lambda})e^{il\theta},\nonumber
\end{eqnarray}
the integration in $\theta'$  in equation (\ref{eq:deltaH_general}) can be carried out by transforming into transit harmonics, and one obtains
\begin{eqnarray}
\delta H_s=i\omega\hat{S}\sum_p i^p J_p(\hat{\Lambda})e^{ip\theta}\sum_l \frac{(-i)^lJ_l(\hat{\Lambda})e^{il\theta}}{\omega-l\omega_{tr}}.\label{eq:Hi_harmonic}
\end{eqnarray}
Here, $\hat{\Lambda}\equiv \hat{\omega}_d/\omega_{tr}$  and $\exp{(-i\hat{\Lambda}\cos\theta)}$ is the ``pullback" (coordinate transformation) from  drift orbit center to particle guiding center coordinates. The resonance condition is $\omega-l\omega_{tr}=0$, with $l$ being  integer,   and resonant particles satisfying $|v_{\parallel,res}/v_{ti}|\sim O(q/l)$  due to the GAM/EGAM frequency ordering. The subscript ``res" denotes resonant particles.   Furthermore, the ``population" of particles for each transit resonances is proportional to $J^2_l(\hat{\Lambda})\partial_E F_0|_{v_{\parallel,res}}$.  Noting that $\hat{\Lambda}_{res}\sim k_r\rho_i q^2/l$ and the properties of Bessel functions,  one can truncate the summation in equation (\ref{eq:Hi_harmonic}) at finite $l$ \cite{FHintonPPCF1999,HSugamaJPP2006}  in the small drift orbit limit with $k_r\rho_i q^2\ll1$.  GAM collisionless damping   due to the primary transit resonance ($|\omega|=|\omega_{tr}|$) only was   investigated in Ref. \citenum{FHintonPPCF1999}. It was shown by Sugama et al \cite{HSugamaJPP2006} that, for increasing $k_r\rho_{ti}q^2$,   GAM collisionless damping can be significantly enhanced by the increasing weight of higher order transit resonances due to the finite orbit width effect;  and the analytical expression including $|\omega|=2|\omega_{tr}|$ resonance was derived. By further increasing $\hat{\Lambda}_{res}$  due to larger  $k_r$ or $q$, however, more and more transit resonances are needed for the accurate description of GAM collisionless damping \cite{XXuPRL2008}, and the analytical expression is very difficult to obtain due to the non-trivial task of  summing up all the transit resonances.

An alternative approach was developed in Ref. \cite{FZoncaEPL2008}, to derive the analytical expression of GAM collisionless damping  rate in the short wavelength limit ($k_r\rho_iq^2\gg1$), with all the transit   resonances taken into account.   Here, we  will first show that, the perturbed distribution function for resonant particles derived  in Refs. \cite{FZoncaEPL2008,ZQiuPPCF2009} are equivalent to the general solution of equations (\ref{eq:deltaH_general}) or (\ref{eq:Hi_harmonic}) in the proper  limit, and then
briefly summarize the main idea  of this approach \cite{FZoncaEPL2008}; while interested readers   may refer to Ref. \cite{ZQiuPPCF2009} for the detailed derivation. 

In the large orbit limit, equation (\ref{eq:deltaH_general}) can be expanded using the smallness parameter $1/\dot{\psi}$, with  $|\dot{\psi}|\sim |\hat{\omega}_d/\omega_{tr}|\gg1$ in the large orbit limit and having denoted derivation of $\psi(\theta)$ with respect to $\theta$ as $\dot \psi$ for brevity.  Noting that
\begin{eqnarray}
\int^{\theta}_{-\infty}  e^{\psi(\theta')}d\theta'&=&\frac{e^{\psi}}{\dot{\psi}}-\frac{e^{\psi}}{\dot{\psi}}\frac{\partial}{\partial\theta}\frac{1}{\dot{\psi}}+\frac{e^{\psi}}{\dot{\psi}}\frac{\partial}{\partial\theta}\left(\frac{1}{\dot{\psi}}\frac{\partial}{\partial\theta}\frac{1}{\dot{\psi}}\right)\nonumber\\
&-& \int^{\theta}_{-\infty}e^{\psi(\theta')}\frac{\partial}{\partial\theta'}\left(\frac{1}{\dot{\psi}}\frac{\partial}{\partial\theta'}\left(\frac{1}{\dot{\psi}}\frac{\partial}{\partial\theta'}\frac{1}{\dot{\psi}}\right)\right)d\theta',\nonumber
\end{eqnarray}
one then has
\begin{eqnarray}
\delta H_s&=&\frac{\omega}{\omega_{tr}}\hat{S}\left[\frac{1}{\dot{\psi}} -\frac{1}{2}\frac{\partial}{\partial\theta}\left(\frac{1}{\dot{\psi}}\right)^2+\frac{1}{2\dot{\psi}}\frac{\partial^2}{\partial \theta^2}\left(\frac{1}{\dot{\psi}}\right)^2\right.\nonumber\\
&&\hspace{11em}\left.+ O(\dot{\psi}^{(-4)})\right].\label{eq:deltaH_asym}
\end{eqnarray}

Noting that $\dot{\psi}=-i (\omega-\hat{\omega}_d\sin\theta)/\omega_{tr}$, the three terms in the square bracket of  equation (\ref{eq:deltaH_asym}) corresponds,  respectively,  to $\delta H^{(0)}_{res}$, $\delta H^{(1)}_{res}$ and $\delta H^{(2)}_{res}$ in equations (16), (21) and (23) of Ref. \cite{ZQiuPPCF2009}, in the $T_e/T_i\ll1$ limit assumed here. Thus, the $\delta H_{res}$'s in Ref. \cite{ZQiuPPCF2009} are equivalent to the general solution of equation (\ref{eq:Hi_harmonic}) by summing up all the transit harmonics, and the underlying wave-paricle interactions  in the short wavelength limit are indeed through transit resonances, as pointed out in Ref. \cite{ZQiuPPCF2009}.  The first term in equation (\ref{eq:deltaH_asym}) corresponds to the perturbed resonant particle distribution  function in  the  $q\rightarrow\infty$ limit;   the third term gives the $O(1/q^2)$ corrections while the second term vanishes in the surface average.

Since we are interested in the collisionless damping due to thermal ion contribution, a single thermal ion species  with Maxwellian distribution function can be assumed, and the GAM dielectric function is derived from the surface averaged quasi-neutrality condition
\begin{eqnarray}
D_G\equiv \left.\left\langle  -\frac{e}{T_i}F_0\overline{\delta\phi}_G+J_G\overline{\delta H_i} \right\rangle\right/\left(\frac{e}{T_i}n_0\overline{\delta\phi}_G\right).\nonumber
\end{eqnarray}
The    imaginary part of $D_G$ due to resonant particle contribution,  to the leading order,   is   then
\begin{eqnarray}
D^{(0)}_i=\mathbb{I}{\rm m}\left\langle \frac{F_0}{n_0}J^2_G\omega\int \frac{d\theta}{2\pi}\frac{1}{\omega-\omega_d}\right\rangle.\label{eq:GAM_Di}
\end{eqnarray}

We note that, even though in equation (\ref{eq:GAM_Di})  the anti-Herimitian part comes from the imaginary part of $1/(\omega-\omega_d)$, the  underlying interaction is not a ``drift resonance"  \cite{MSasakiPoP2016}, since $\omega_d\propto\sin\theta$ is temporally fast varying and the effective  energy exchange is due to transit resonances as shown in equation (\ref{eq:Hi_harmonic}). The  surface average is then carried out by expanding $\omega_d$ round  $\theta=\pm \pi/2$ where $|\omega_d|$ is maximized and the integration in $\theta$ is performed by the method of steepest descent. Again, readers  interested in the details of the algebra can consult Ref. \citenum{ZQiuPPCF2009}.  Here, we will briefly summarize   the main ideas underlying   the derivation: 
\begin{itemize}
\item[1]   considering the wave-particle interaction on the time scale of $|\omega_d|^{-1}$, which is much shorter than the transit time $|\omega_{tr}|^{-1}$ in the large orbit limit, corresponds to the inclusion of  a broad spectrum in frequency, i.e., all the transit harmonics are taken into account; 
\item[2] for  resonant particles,  the dominant energy exchange with GAM is captured noting that the wave- particle energy exchange is caused by the acceleration in the radial direction associated with the radial magnetic drift, i.e., $\dot{E}=(e/m)\mathbf{V}_d\cdot\delta\mathbf{E}_r$, which maximises around $|\theta|=\pi/2$. Here, $V_d\equiv (v^2_{\perp}/2+v^2_{\parallel})\sin\theta \mathbf{e}_r/(\Omega_i R_0)$ is the radial component of magnetic drift velocity. 
\item[3]Noting again that $\omega_d\propto\sin\theta$ is maximized around $|\theta|=\pi/2$, ions with lower energy and thus,  proportionally (exponentially for a Maxwellian distribution with typical parameters) larger population, will contribute to the resonance. 
\end{itemize}

As a further application, EGAM excitation by EPs in the large magnetic drift  orbit limit will be investigated; which is part of  the motivation of this communication.  To focus on the wave-particle resonance in the short wavelength limit considering the effect of finite magnetic drift orbit averaging, we take $T_e/T_i\ll1$ and further neglect the finite Larmor radius effect of EPs. Thus, the leading order EP response to GAM can be derived as
\begin{eqnarray}
\delta H_h=-\frac{e}{m}\partial_E F_{0h}\overline{\delta\phi}_G\frac{\omega}{\omega-\omega_d},\nonumber
\end{eqnarray}
and the linear dispersion relation of EGAM can  be obtained from the quasi-neutrality condition
\begin{eqnarray}
\hat{\mathscr{E}}_{EGAM}\equiv\left.\left(\overline{\delta n_i}+\overline{\delta n_h}\right)\right/\left(en_0\overline{\delta\phi}_G/T_i\right).\nonumber
\end{eqnarray}

As the expression of  thermal ion density perturbation  can be found in Ref. \citenum{FZoncaEPL2008},   we will focus on the EP density perturbation, 
\begin{eqnarray}
\overline{\delta n_h}
&=&  -\frac{e}{m}B_0\sum_{\sigma=\pm1}\int \frac{E dE d\Lambda}{|v_{\parallel}|}\int  d\theta \frac{\partial F_{0h}}{\partial E}\overline{\delta\phi}_G\frac{\omega_d}{\omega-\omega_d}.\nonumber
\end{eqnarray}
Here, $\Lambda = \mu/E$ is the usual definition of the particle pitch angle in velocity space, with $\mu=v_\perp^2/(2B)$  the magnetic moment.  Noting that $\omega_d=\hat{\omega}_d\sin\theta$ maximizes  at $\theta\simeq \pi/2$, the contribution around $\theta\simeq \pm\pi/2$ dominates where wave-particle power exchange maximizes.  Taking $x=\theta-\mbox{sign}(\theta) \pi/2$, one then has 
\begin{eqnarray}
\int d\theta\frac{\omega_d}{\omega-\omega_d}&=&-2\pi+\omega\int^{\infty}_{-\infty} dx \frac{1}{\omega-\hat{\omega}_d(1-x^2/2)}\nonumber\\
&=&-2\pi\frac{i}{\sqrt{(2\hat{\omega}_d/\omega)(\hat{\omega}_d/\omega-1)}}.\label{eq:surface_averaged}
\end{eqnarray}
In equation (\ref{eq:surface_averaged}), the contribution of non-resonant adiabatic particle response is neglected, and the perturbed EP density is then
\begin{eqnarray}
\overline{\delta n_h}&=&2\pi i B_0\frac{e}{m}\overline{\delta\phi}_G\nonumber\\
&\times& \sum_{\sigma=\pm1}\int\frac{EdEd\Lambda}{|v_{\parallel}|}\frac{\partial_EF_{0h}}{\sqrt{(2\hat{\omega}_d/\omega)(\hat{\omega}_d/\omega-1)}}.
\end{eqnarray}
Taking a single-pitch angle slowing down EP distribution function \cite{ZQiuPPCF2010} as that for neutral beam injection, i.e., 
$F_{0h}=c_0\delta(\Lambda-\Lambda_0)H_E$, with $c_0=n_b\sqrt{2(1-\Lambda_0B_0)}/(4\pi B_0\ln(E_b/E_c))$, $n_b$ is the density of the EP beam,  $E_b$ and $E_c$ being respectively the EP birth and critical energies,   $\delta(x)$ is the Dirac delta function, 
and  $H_E=1/(E^{3/2}+E^{3/2}_c)\Theta(1-E/E_b)$ with $\Theta(1-E/E_b)$ being the Heaviside step function.  The integration in velocity space can then be carried out, and yields the short wavelength EGAM  dispersion relation: 
\begin{eqnarray}
\hat{b}_i\left(-1+\frac{\omega^2_G}{\omega^2}\right)+\Delta_f&+& i n_b\left[\frac{-2+3\Lambda_0B_0}{1-\Lambda_0B_0}\frac{\Omega_b}{\omega}\sqrt{\frac{\Omega_b}{\omega}-1}\right.\nonumber\\
&&\left.-\Lambda_0B_0\frac{(\omega/\Omega_b)^{1/2}}{\sqrt{\Omega_b/\omega-1}}\right]=0,\label{eq:DR_final}
\end{eqnarray}
with $\Delta_f$ being the non-resonant EP contribution
\begin{eqnarray}
\Delta_f=n_b\left[\frac{-2+3\Lambda_0B_0}{1-\Lambda_0B}\frac{\Omega_b}{\omega}+\Lambda_0B_0\left(\frac{\omega}{\Omega_b}\right)^{1/2}\left(\frac{E_b}{E_c}\right)^{3/2}\right],\nonumber
\end{eqnarray}
$\Omega_b\equiv\hat{\omega}_d(E=E_b)$, $\hat{b}_i=k^2_r\rho^2_{ti}/2$, and $\omega_G\simeq\sqrt{7/4+\tau}v_{ti}/R_0$ is the GAM frequency.

\begin{figure}
\includegraphics[width=9cm]{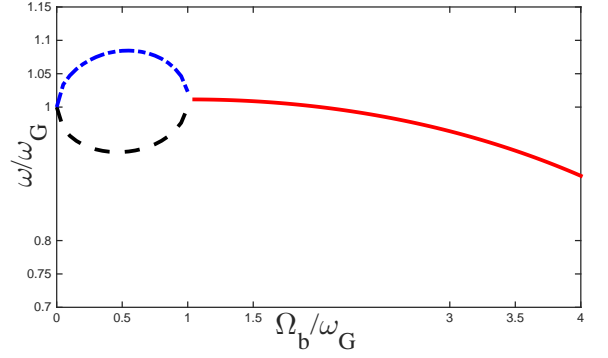}
\caption{Real frequency v.s. $\Omega_b/\omega_G$}\label{fig:RF}
\end{figure}

The first term in the square bracket of  equation (\ref{eq:DR_final}) ($\propto\sqrt{\Omega_b/\omega-1}$) could be the destabilizing term depending on the value of $\Lambda_0B_0$, while the second term ($\propto(\Omega_b/\omega-1)^{-1/2}$) is  stabilizing. 
As a result, EGAM excitation in the large orbit limit requires, first,
\begin{eqnarray}
\Lambda_0B_0>2/3,\label{eq:drive}
\end{eqnarray}
for the first term of EP contribution in equation (\ref{eq:DR_final}) to be destabilizing; and second,  $\Omega_b/\omega$  being sufficiently large for the short wavelength EGAM to be unstable.

\begin{figure}
\includegraphics[width=9cm]{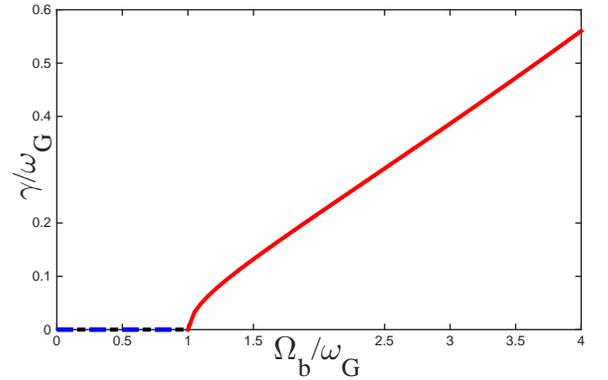}
\caption{Growth rate v.s. $\Omega_b/\omega_G$}\label{fig:GR}
\end{figure}

The dispersion relation is solved numerically as a function of $\Omega_b/\omega_G$. Note that from our previous analysis,  the drive due to wave-particle resonance exists only for $|\Omega_b/\omega|>1$; and that the destabilizing term increases with $|\Omega_b/\omega|$ while the stabilizing term decreases with $|\Omega_b/\omega|$, so the destabilizing term is neglected, which gives negligible contribution for  $|\Omega_b/\omega|>1$ where wave-particle energy exchange exists. The other parameters are taken as follows: $n_b/\hat{b}_i=0.3$, $\Delta_f=0$, and the obtained short wavelength EGAM real frequency and growth rate dependences on $\Omega_b/\omega_G$ are shown in Figs. \ref{fig:RF} and \ref{fig:GR}, respectively.  It is shown that, the    EGAM real frequency decreases slightly with increasing $\Omega_b/\omega_G$, and the unstable EGAM frequency is always smaller than local GAM frequency. 
On the other hand, as the EGAM is unstable for $\Omega_b/\omega_G>1$, the growth rate increases  with   $\Omega_b$.  For $\Omega_b/\omega_G$ significantly larger than unity, the growth rate increases almost linearly with $\Omega_b/\omega_G$, and thus, $E_b$, as is clearly seen from  the destabilizing term of equation (\ref{eq:DR_final}).  This is due to the increasingly  dense high  order transit resonances associated with increasing $E_b$. Whereas,  in the long wavelength limit, the growth rate will be  peaked when the EP parallel velocity at birth energy satisfies a certain harmonic resonance, similar to the case for GAM Landau damping discussed in  Ref. \cite{HSugamaJPP2006}.

In conclusion,  the underlying physics picture of wave-particle resonances at short wavelength is clarified, taking short wavelength  GAM collisionless damping  as an example.  Assuming large aspect ratio tokamak  and well circulating particles, the   ion response to GAM is derived from linear gyrokinetic equation by integration  along unperturbed guiding-center orbit. The general solution is then obtained by expansion into transit harmonics, with the ``population" of resonant particles to each transit harmonic  proportional to $J^2_l(\hat{\Lambda}_{res})\partial_E F_0|_{v_{\parallel,res}}$.  As a result,  to obtain the GAM collisionless damping in the short wavelength limit, all the transit resonances must be kept. It is then shown that, the result obtained in Ref. \citenum{FZoncaEPL2008} based on large orbit width expansion, is equivalent to the general solution up to $O(1/(k_r\rho_{ti}q^2))$; and the underlying physics  for wave-particle interactions at short wavelength consists indeed  in the summation of all the transit resonances.

As a further application, the EGAM excitation at short wavelengths is also investigated, and the analytical dispersion relation is derived assuming a single pitch angle slowing down EP distribution function. Our results indicates that  the   short wavelength EGAM dispersion relation depends algebraically on the EP characteristic frequency, instead of the logarithmic dependence characterizing the long wavelength limit, which  is typical for a slowing down EP distribution function. The short wavelength EGAM is unstable for $\Omega_b>\omega_G$, and $\Lambda_0B_0>2/3$. For $\Omega_b$ significantly larger than GAM frequency, the short wavelength EGAM growth rate is proportional to $\Omega_b$, and thus, EP birth energy $E_b$ due to the increasingly denser high order transit resonances as $\Omega_b\gg\omega_G$.

This work is supported by     US DoE GRANT,  the National Magnet Confinement Fusion Research Program under Grants Nos.  2013GB104004  and   2013GB111004,
the National Science Foundation of China under grant Nos.  11575157  and 11235009,  Fundamental Research Fund for Chinese Central Universities under Grant No. 2017FZA3004  and   EUROfusion Consortium
under grant agreement No. 633053.


\begin{thebibliography}{14}
\expandafter\ifx\csname natexlab\endcsname\relax\def\natexlab#1{#1}\fi
\expandafter\ifx\csname bibnamefont\endcsname\relax
  \def\bibnamefont#1{#1}\fi
\expandafter\ifx\csname bibfnamefont\endcsname\relax
  \def\bibfnamefont#1{#1}\fi
\expandafter\ifx\csname citenamefont\endcsname\relax
  \def\citenamefont#1{#1}\fi
\expandafter\ifx\csname url\endcsname\relax
  \def\url#1{\texttt{#1}}\fi
\expandafter\ifx\csname urlprefix\endcsname\relax\def\urlprefix{URL }\fi
\providecommand{\bibinfo}[2]{#2}
\providecommand{\eprint}[2][]{\url{#2}}

\bibitem[{\citenamefont{Winsor et~al.}(1968)\citenamefont{Winsor, Johnson, and
  Dawson}}]{NWinsorPoF1968}
\bibinfo{author}{\bibfnamefont{N.}~\bibnamefont{Winsor}},
  \bibinfo{author}{\bibfnamefont{J.~L.} \bibnamefont{Johnson}},
  \bibnamefont{and} \bibinfo{author}{\bibfnamefont{J.~M.}
  \bibnamefont{Dawson}}, \bibinfo{journal}{Physics of Fluids}
  \textbf{\bibinfo{volume}{11}}, \bibinfo{pages}{2448} (\bibinfo{year}{1968}).

\bibitem[{\citenamefont{Sasaki et~al.}(2016)\citenamefont{Sasaki, Kasuya, Itoh,
  Hallatschek, Lesur, Kosuga, and Itoh}}]{MSasakiPoP2016}
\bibinfo{author}{\bibfnamefont{M.}~\bibnamefont{Sasaki}},
  \bibinfo{author}{\bibfnamefont{N.}~\bibnamefont{Kasuya}},
  \bibinfo{author}{\bibfnamefont{K.}~\bibnamefont{Itoh}},
  \bibinfo{author}{\bibfnamefont{K.}~\bibnamefont{Hallatschek}},
  \bibinfo{author}{\bibfnamefont{M.}~\bibnamefont{Lesur}},
  \bibinfo{author}{\bibfnamefont{Y.}~\bibnamefont{Kosuga}}, \bibnamefont{and}
  \bibinfo{author}{\bibfnamefont{S.-I.} \bibnamefont{Itoh}},
  \bibinfo{journal}{Physics of Plasmas} \textbf{\bibinfo{volume}{23}},
  \bibinfo{pages}{102501} (\bibinfo{year}{2016}).

\bibitem[{\citenamefont{Zonca and Chen}(2008)}]{FZoncaEPL2008}
\bibinfo{author}{\bibfnamefont{F.}~\bibnamefont{Zonca}} \bibnamefont{and}
  \bibinfo{author}{\bibfnamefont{L.}~\bibnamefont{Chen}},
  \bibinfo{journal}{Europhys. Lett.} \textbf{\bibinfo{volume}{83}},
  \bibinfo{pages}{35001} (\bibinfo{year}{2008}).

\bibitem[{\citenamefont{Qiu et~al.}(2009)\citenamefont{Qiu, Chen, and
  Zonca}}]{ZQiuPPCF2009}
\bibinfo{author}{\bibfnamefont{Z.}~\bibnamefont{Qiu}},
  \bibinfo{author}{\bibfnamefont{L.}~\bibnamefont{Chen}}, \bibnamefont{and}
  \bibinfo{author}{\bibfnamefont{F.}~\bibnamefont{Zonca}},
  \bibinfo{journal}{Plasma Physics and Controlled Fusion}
  \textbf{\bibinfo{volume}{51}}, \bibinfo{pages}{012001}
  (\bibinfo{year}{2009}).

\bibitem[{\citenamefont{Zhang et~al.}(2008)\citenamefont{Zhang, Lin, Chen
  et~al.}}]{WZhangPRL2008}
\bibinfo{author}{\bibfnamefont{W.}~\bibnamefont{Zhang}},
  \bibinfo{author}{\bibfnamefont{Z.}~\bibnamefont{Lin}},
  \bibinfo{author}{\bibfnamefont{L.}~\bibnamefont{Chen}},
  \bibinfo{journal}{Physical review letters} \textbf{\bibinfo{volume}{101}},
  \bibinfo{pages}{095001} (\bibinfo{year}{2008}).

\bibitem[{\citenamefont{Feng et~al.}(2013)\citenamefont{Feng, Qiu, and
  Sheng}}]{ZFengPoP2013}
\bibinfo{author}{\bibfnamefont{Z.}~\bibnamefont{Feng}},
  \bibinfo{author}{\bibfnamefont{Z.}~\bibnamefont{Qiu}}, \bibnamefont{and}
  \bibinfo{author}{\bibfnamefont{Z.}~\bibnamefont{Sheng}},
  \bibinfo{journal}{Physics of Plasmas (1994-present)}
  \textbf{\bibinfo{volume}{20}}, \bibinfo{eid}{122309} (\bibinfo{year}{2013}).

\bibitem[{\citenamefont{Nazikian et~al.}(2008)\citenamefont{Nazikian, Fu, and
  Austin}}]{RNazikianPRL2008}
\bibinfo{author}{\bibfnamefont{R.}~\bibnamefont{Nazikian}},
  \bibinfo{author}{\bibfnamefont{G.}~\bibnamefont{Fu}}, 
   \bibinfo{author}{\bibfnamefont{M.}~\bibnamefont{Austin}},
 \bibinfo{author}{\bibfnamefont{H.}~\bibnamefont{Berk}},
  \bibinfo{author}{\bibfnamefont{R.}~\bibnamefont{Budny}},
 \bibinfo{author}{\bibfnamefont{N.}~\bibnamefont{Gorelenkov}},
 \bibinfo{author}{\bibfnamefont{W.}~\bibnamefont{Heidbrink}},  
 \bibinfo{author}{\bibfnamefont{C.}~\bibnamefont{Holcomb}} , 
 \bibinfo{author}{\bibfnamefont{G.}~\bibnamefont{Kramer}}, 
  \bibinfo{author}{\bibfnamefont{G.}~\bibnamefont{McKee}},  
   \bibinfo{author}{\bibfnamefont{M.}~\bibnamefont{Makowski}},
    \bibinfo{author}{\bibfnamefont{W.}~\bibnamefont{Solomon}},  
     \bibinfo{author}{\bibfnamefont{M.}~\bibnamefont{Shafer}},  
      \bibinfo{author}{\bibfnamefont{E.}~\bibnamefont{Strait}},
      \bibnamefont{and} 
\bibinfo{author}{\bibfnamefont{M.}~\bibnamefont{Van Zeeland}},
  \bibinfo{journal}{Phys. Rev. Lett.} \textbf{\bibinfo{volume}{101}},
  \bibinfo{pages}{185001} (\bibinfo{year}{2008}).
  
  

\bibitem[{\citenamefont{Fu}(2008)}]{GFuPRL2008}
\bibinfo{author}{\bibfnamefont{G.}~\bibnamefont{Fu}}, \bibinfo{journal}{Phys.
  Rev. Lett.} \textbf{\bibinfo{volume}{101}}, \bibinfo{pages}{185002}
  (\bibinfo{year}{2008}).

\bibitem[{\citenamefont{Qiu et~al.}(2010)\citenamefont{Qiu, Zonca, and
  Chen}}]{ZQiuPPCF2010}
\bibinfo{author}{\bibfnamefont{Z.}~\bibnamefont{Qiu}},
  \bibinfo{author}{\bibfnamefont{F.}~\bibnamefont{Zonca}}, \bibnamefont{and}
  \bibinfo{author}{\bibfnamefont{L.}~\bibnamefont{Chen}},
  \bibinfo{journal}{Plasma Phys. Control. Fusion} \textbf{\bibinfo{volume}{52}}
  (\bibinfo{year}{2010}).

\bibitem[{\citenamefont{Rutherford and Frieman}(1968)}]{PRutherfordPoF1968}
\bibinfo{author}{\bibfnamefont{P.}~\bibnamefont{Rutherford}} \bibnamefont{and}
  \bibinfo{author}{\bibfnamefont{E.}~\bibnamefont{Frieman}},
  \bibinfo{journal}{Physics of Fluids} \textbf{\bibinfo{volume}{11}},
  \bibinfo{pages}{569} (\bibinfo{year}{1968}).

\bibitem[{\citenamefont{Taylor and Hastie}(1968)}]{JTaylorPP1968}
\bibinfo{author}{\bibfnamefont{J.}~\bibnamefont{Taylor}} \bibnamefont{and}
  \bibinfo{author}{\bibfnamefont{R.}~\bibnamefont{Hastie}},
  \bibinfo{journal}{Plasma Physics} \textbf{\bibinfo{volume}{10}},
  \bibinfo{pages}{479} (\bibinfo{year}{1968}).

\bibitem[{\citenamefont{Hinton and M.}(1999)}]{FHintonPPCF1999}
\bibinfo{author}{\bibfnamefont{F.}~\bibnamefont{Hinton}} \bibnamefont{and}
  \bibinfo{author}{\bibfnamefont{R.}~\bibnamefont{M.}},
  \bibinfo{journal}{Plasma Physics and Controlled Fusion}
  \textbf{\bibinfo{volume}{41}}, \bibinfo{pages}{A653} (\bibinfo{year}{1999}).

\bibitem[{\citenamefont{Sugama and Watanabe}(2006)}]{HSugamaJPP2006}
\bibinfo{author}{\bibfnamefont{H.}~\bibnamefont{Sugama}} \bibnamefont{and}
  \bibinfo{author}{\bibfnamefont{T.-H.} \bibnamefont{Watanabe}},
  \bibinfo{journal}{Journal of plasma physics} \textbf{\bibinfo{volume}{72}},
  \bibinfo{pages}{825} (\bibinfo{year}{2006}).

\bibitem[{\citenamefont{Xu et~al.}(2008)\citenamefont{Xu, Xiong, Gao, Nevins,
  and McKee}}]{XXuPRL2008}
\bibinfo{author}{\bibfnamefont{X.}~\bibnamefont{Xu}},
  \bibinfo{author}{\bibfnamefont{Z.}~\bibnamefont{Xiong}},
  \bibinfo{author}{\bibfnamefont{Z.}~\bibnamefont{Gao}},
  \bibinfo{author}{\bibfnamefont{W.}~\bibnamefont{Nevins}}, \bibnamefont{and}
  \bibinfo{author}{\bibfnamefont{G.}~\bibnamefont{McKee}},
  \bibinfo{journal}{Physical review letters} \textbf{\bibinfo{volume}{100}},
  \bibinfo{pages}{215001} (\bibinfo{year}{2008}).

\end{thebibliography}
\end{document}